\documentclass[8.5pt,twoside,twocolumn]{article}
\usepackage[super,sort&compress,comma]{natbib} 
\usepackage{times,mathptmx}
\usepackage{sectsty}
\usepackage{balance}
\usepackage{color,bm,epsfig,graphics,amssymb,amsmath,subeqnarray,setspace,graphicx,amsthm,epstopdf,subfigure,color} 

\usepackage{bm,epsfig,graphics,amssymb,amsmath,subeqnarray,setspace,graphicx,amsthm,epstopdf,subfigure,color}
\usepackage{epstopdf,psfrag, graphicx,color}
\usepackage{widetext}

\usepackage{graphicx} %eps figures can be used instead
\usepackage{lastpage}
\usepackage[format=plain,justification=justified,singlelinecheck=false,font=small,labelfont=bf,labelsep=space]{caption} %justification=raggedright
\usepackage{fancyhdr}
\usepackage{bm,epsfig,graphics,amssymb,amsmath,subeqnarray,setspace,graphicx,amsthm,epstopdf,subfigure,color}
\usepackage{epstopdf, epsf,psfrag, graphicx,color}
\usepackage{float}
\usepackage{multicol}

\usepackage{lipsum}
\usepackage{mathtools}
\usepackage{cuted}

\floatstyle{plain}
\newfloat{twocolequfloat}{b}{zzz}
\floatname{twocolequfloat}{Equation}
\usepackage{mathptmx}
\DeclareSymbolFont{greekletters}{OML}{cmr}{m}{it}
\DeclareMathSymbol{\varrho}{\mathalpha}{greekletters}{"25}

\definecolor{DarkBlue}{rgb}{0,0,.8}

\begin{document}
%\begin{multicols}{2}%

\thispagestyle{plain}
\fancypagestyle{plain}{
%\fancyhead[L]{\includegraphics[height=8pt]{headers/LH}}
%\fancyhead[C]{\hspace{-1cm}\includegraphics[height=20pt]{headers/CH}}
%\fancyhead[R]{\includegraphics[height=10pt]{headers/RH}\vspace{-0.2cm}}
\renewcommand{\headrulewidth}{1pt}}
\renewcommand{\thefootnote}{\fnsymbol{footnote}}
\renewcommand\footnoterule{\vspace*{1pt}% 
\hrule width 3.4in height 0.4pt \vspace*{5pt}} 
\setcounter{secnumdepth}{5}

\makeatletter 
\def\subsubsection{\@startsection{subsubsection}{3}{10pt}{-1.25ex plus -1ex minus -.1ex}{0ex plus 0ex}{\normalsize\bf}} 
\def\paragraph{\@startsection{paragraph}{4}{10pt}{-1.25ex plus -1ex minus -.1ex}{0ex plus 0ex}{\normalsize\textit}} 
\renewcommand\@biblabel[1]{#1}            
\renewcommand\@makefntext[1]% 
{\noindent\makebox[0pt][r]{\@thefnmark\,}#1}
\makeatother 
\renewcommand{\figurename}{\small{Fig.}~}
\sectionfont{\large}
\subsectionfont{\normalsize}

\fancyfoot{}
%\fancyfoot[LO,RE]{\vspace{-7pt}\includegraphics[height=9pt]{headers/LF}}
%\fancyfoot[CO]{\vspace{-7.2pt}\hspace{12.2cm}\includegraphics{headers/RF}}
%\fancyfoot[CE]{\vspace{-7.5pt}\hspace{-13.5cm}\includegraphics{headers/RF}}
\fancyfoot[RO]{\footnotesize{\sffamily{1--\pageref{LastPage} ~\textbar  \hspace{2pt}\thepage}}}
\fancyfoot[LE]{\footnotesize{\sffamily{\thepage~\textbar\hspace{3.45cm} 1--\pageref{LastPage}}}}
\fancyhead{}
\renewcommand{\headrulewidth}{1pt} 
\renewcommand{\footrulewidth}{1pt}
\setlength{\arrayrulewidth}{1pt}
\setlength{\columnsep}{6.5mm}
\setlength\bibsep{1pt}

\twocolumn[
  \begin{@twocolumnfalse}
\noindent\LARGE{\textbf{Stability of the interface of an isotropic active fluid}}
\vspace{0.6cm}

\noindent\large{\textbf{Harsh Soni\textit{$^{a}$}, Wan Luo\textit{$^{a}$}, Robert A. Pelcovits\textit{$^{b}$}, and Thomas Powers\textit{$^{a,b}$}}}\vspace{0.5cm}
%Please note that \ast indicates the corresponding author(s) but no footnote text is required. 
%\noindent\textit{\small{\textbf{\text{Re}ceived Xth XXXXXXXXXX 20XX, Accepted Xth XXXXXXXXX 20XX\newline
%First published on the web Xth XXXXXXXXXX 20XX}}}
\\
\noindent{\today}

%\noindent \textbf{\small{DOI: }}
\vspace{0.6cm}
%Please do not change this text.
 %less than 250 words

\noindent\normalsize{We study the linear stability of an isotropic active fluid in three different geometries: a film of active fluid on a rigid substrate, a cylindrical thread of fluid, and a spherical fluid droplet. The active fluid is modeled by the hydrodynamic theory of an active nematic liquid crystal in the isotropic phase. In each geometry, we calculate the growth rate of sinusoidal modes of deformation of the interface. There are two distinct branches of growth rates; at long wavelength, one corresponds to the deformation of the interface, and one corresponds to the evolution of the liquid crystalline degrees of freedom. The passive cases of the film and the spherical droplet are always stable. For these geometries, a sufficiently large activity leads to instability. Activity also leads to propagating damped or growing modes. The passive cylindrical thread is unstable for perturbations with wavelength longer than the circumference. A sufficiently large activity can make any wavelength unstable, and again leads to propagating damped or growing modes.   }

\vspace{0.5cm}
 \end{@twocolumnfalse}
]

\footnotetext{\textit{$^{a}$~School of Engineering, Brown University, Providence, RI 02912, USA. }}
\footnotetext{\textit{$^{b}$~Department of Physics, Brown University, Providence, RI 02912, USA. }}

%, their interfaces can also be destabilized by their own dynamics The interface surface energies affacts the instability.
\section{Introduction}
% and insensibilities of the droplets and thin films

Active fluids are energized locally by motorized microscopic active particles such as kinesin-driven microtubules~\cite{Sanchez2012} and myosin-actin complexes~\cite{Mizuno370}. Therefore, their dynamics occur out of thermal equilibrium~\cite{ramaswamy2017,srimrmp}. Hydrodynamic instabilities of both polar and nematic active fluids have been studied using hydrodynamic theories and simulations for bulk~\cite{sriram2002,SaintillanShelley2008,PahlavanSaintillan2011} as well as for confined fluids~\cite{sriramThinfilm2009,Maitra2018,WoodhouseGoldstein2012,Giomi_etal2012,Norton2018, Theillard2017, Edwards2009}. In this paper, we consider the  instabilities of active nematic fluids in the isotropic phase confined by an interface.   The damping of a capillary wave on a flat interface between two passive viscous fluids is well-understood~\cite{Lamb1994}. Likewise, theoretical studies of interfacial instabilities like the Rayleigh-Plateau capillary instability and Rayleigh-Taylor interface instability have been carried out for passive fluids~\cite{Tomotika1935} including complex fluids such as polymer solutions~\cite{Stanley1965,goldin1969} and liquid crystals~\cite{Cheong2001,Cheong2002}. Less work has been done on interfacial instabilities in active fluids. It is natural to expect that the instabilities that occur in bulk active fluids can destabilize an otherwise stable interface, or make an already unstable interface more unstable. Work to date includes a study by Yang and Wang~\cite{softmatter2014} of the Rayleigh-Plateau capillary instability of a thread of active polar fluid in the ordered state surrounded by a passive Newtonian fluid, a study by Whitfield and Hawkins~\cite{2016njpactivedrop} of the instability of an spherical droplet of active polar fluid in the ordered state, and an analysis by Gao and Li~\cite{activedroplet2017prl} of a self-driven droplet of an active nematic fluid. Also, Patteson \textit{et al.}~\cite{Patteson2018} studied the propagation of active-passive interfaces in bacterial swarms. Recently, Maitra \textit{et al.}~\cite{sriramMembrane2014} explored the dynamics of an active membrane in an active polar medium,  Mietke \textit{et al.} studied the instabilities of an active membrane in a passive fluid~\cite{MietkeJulicherSbalzarini2019}, and V. Soni \textit{et al.} studied the surface dynamics of an active colloidal chiral fluid~\cite{VSoni_etal2018}. Here, we focus on linear stability analyses of  active nematic fluids in the isotropic phase in with flat, cylindrical, or spherical interfaces.  Our focus on the isotropic phase is motivated by recent experiments on active matter that show large regions in which the nematic order is small~\cite{Wu_etal2017}. Our work is distinct from the other theoretical work just mentioned on interfacial instabilities in active fluids because we consider the active nematic to initially be in the isotropic state instead of the ordered state. 
 \begin{figure}[t]
	\includegraphics[width=%0.45\textwidth
	3.3in]{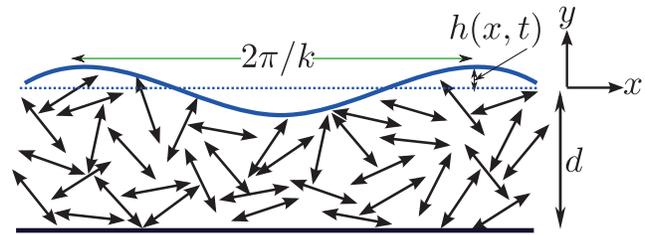}
	%\centering
	\caption{(Color online.) %Schematic diagram of a 
		A film of an isotropic active nematic liquid fluid of depth $d$ in its quiescent state.  The double headed arrows are the active nematic molecules. The bottom surface of the film  is in contact with a solid substrate. The top surface is free. The film is  subject to  a  small-amplitude perturbation of wavenumber $k$.}
	\label{schemf}
\end{figure}

We model the isotropic phase of an active nematic fluid by adding activity to de Gennes' hydrodynamic model~\cite{swimming2018,DeGennes1969,degennesbook,deGennes1971} for the isotropic phase of a passive nematic. This model is appropriate for `shakers' rather than `movers' suspended in a liquid. The model shows that in the linear regime the isotropic active nematic fluid behaves like a viscoelastic fluid, with the viscosity and viscoelasticity growing large as the isotropic-nematic transition is approached~\cite{Hatwalne2004}. 
However, our results for the stability of interfaces are qualitatively different from the passive viscoelastic fluid case due to the orientational degrees of freedom.
We work in the limit of low Reynolds number, where viscous effects dominate inertial effects.  
For a passive fluid, deformations of a surface  or spherical surface always relax, whereas a cylindrical thread is unstable to peristaltic deformations of sufficiently long wavelength. When the fluid is active, deformations of the surfaces in all three cases can be unstable. The instability of a bulk active isotropic fluid drives the instability of the flat and spherical surface, and enhances the Rayleigh-Plateau capillary instability of a cylinder.

Our key results are as follows. In the all cases we consider, the coupled dynamics of the interface of the fluid and the nematic order parameter leads to two modes, with damped or growing propagating waves found for a sufficiently large dimensionless activity.
Likewise, surface tension makes it harder for activity to destabilize an active film or an active fluid confined by a spherical interface, as compared to the unconfined case. 
For the cylindrical thread of radius $R$,  harmonic perturbations of wavenumber $k<1/R$ are always unstable, just as in the passive case. %However, their growth rates can be  \textit{reduced} by varying the dimensionless activity. 
Perturbations with $k>1/R$ become unstable above a critical %dimensionless 
activity increasing with $k$ and the surface tension of the interface. 

The remainder of the paper is organized as follows. In section \ref{model}, we introduce a hydrodynamic model for an active nematic fluid in the isotropic state. In section~\ref{membrane}, we use this model to study the linear stability of a film bound by an interface. Next , in section~\ref{thread}, we consider the stability of a thread of active fluid bound by either an interface. Finally, in section~\ref{droplet} analyze the stability of a spherical drop of active fluid. 
We offer concluding remarks in Section \ref{conclusion}.  Section \ref{appen}, the Appendix, contains additional details relevant to Section \ref{membrane}.

\section{Model}\label{model}
The total free energy of an active isotropic nematic fluid with an interface is 
$\mathcal{F}=
\mathcal{F}_n+\mathcal{F}_i$,
where $\mathcal{F}_n$ is the free energy of the nematic fluid, and $\mathcal{F}_i$ is the energy of the interface. 
Denoting the nematic order parameter field by $Q_{\alpha\beta}$, the nematic free energy is 
\begin{equation}
\mathcal{F}_n=\int d^3x \left[ \frac{A}{2} Q_{\alpha\beta}Q_{\alpha\beta}+\frac{B}{3} Q_{\alpha\beta}Q_{\beta\gamma}Q_{\gamma\alpha}+\frac{C}{4}(Q_{\alpha\beta}Q_{\alpha\beta})^2\right],
\label{Fn}
\end{equation}
where we sum over repeated indices $\alpha, \beta,\dots$ which run over the three spatial coordinates.
We consider the isotropic phase, for which $A>0$. In this case, Frank elasticity can be neglected 
 as long as we are not too near the nematic transition. 

The interface energy is given by %~\cite{Ou-YangHelfrich}
\begin{equation}
\mathcal{F}_i=%\frac{\kappa}{2}\int (2 H)^2\mathrm{d}A+
\int\gamma \mathrm{d}S.\label{Fi}
\end{equation}
where $\gamma$ is the interfacial tension and $\mathrm{d}S$ is the element of area.

We use de Gennes' hydrodynamic model~\cite{DeGennes1969,degennesbook,deGennes1971} of the isotropic phase of a passive nematic fluid of uniform concentration, suitably modified~\cite{swimming2018} to account for activity. In terms of  the fluid velocity field $v_\alpha$, the strain rate and the vorticity tensors are given by  $E_{\alpha\beta}=(\partial_\alpha v_\beta+\partial_\beta v_\alpha)/2$ and $\Omega_{\alpha\beta}=(\partial_\alpha v_\beta-\partial_\beta v_\alpha)/2$, respectively, where $\alpha,\beta=x,y,z$. The rate of change $R_{\alpha\beta}$ of the nematic order parameter $Q_{\alpha\beta}$ relative to the local background fluid is defined as
\begin{equation}
R_{\alpha\beta}=\partial_t Q_{\alpha\beta}+\bm{v}\cdot\bm{\nabla}Q_{\alpha\beta}+\Omega_{\alpha\gamma}Q_{\gamma\beta}-Q_{\alpha\gamma}\Omega_{\gamma\beta}.
\end{equation} 
Then, the viscous stress $\sigma^v_{\alpha\beta}$ and equation of motion for the nematic order parameter $Q_{\alpha\beta}$ are given by~\cite{swimming2018}
\begin{eqnarray}
&&\sigma^v_{\alpha\beta}=2\eta E_{\alpha\beta}+2(\mu+\mu_1) R_{\alpha\beta}+a'Q_{\alpha\beta}\label{stressa},\label{sigmaeq}\\
&&\Phi_{\alpha\beta}=2\mu E_{\alpha\beta}+\nu R_{\alpha\beta},\label{Qeq}
\end{eqnarray}
where $\eta$ and $\nu$ are the shear and rotational viscosities, respectively, $\mu$ couples shear and nematic alignment, $a'$ and $\mu_1$ are activity parameters, and $\Phi_{\alpha\beta}$ is the molecular
field defined as $\Phi_{\alpha\beta}\equiv-\delta \mathcal{F}/\delta Q_{\alpha\beta}$. From Eq.~(\ref{Qeq}) we can see that for the case of small shear rate and steady state, the principal axes of the order parameter align with the principal axes of the strain, with the case $\mu<0$ corresponding to the way that prolate particles align in shear, and the case $\mu>0$ corresponding to the way oblate particles align in shear. 
 
In passive fluids, $a'=0$ and $\mu_1=0$. In that case, the Onsager reciprocal relations \cite{onsager} are obeyed and the positive entropy production rate leads to the relation $\eta\nu-2\mu^2>0$. The active term  $aQ_{\alpha\beta}$ appearing in Eq. \eqref{sigmaeq} accounts for the stress due to the force dipoles associated with the active particles~\cite{sriram2002,swimming2018} with $a'>0$ for contractile particles and $a'<0$ for extensile particles. Since time reversal symmetry and the Onsager relations are violated in active fluids when we do not keep track of the chemical reactions in the theory, the active term  $\mu_1 R_{\alpha\beta}$ is allowed in Eq. \eqref{sigmaeq}. Various other approaches to the active matter equations also violate the Onsager reciprocal relation~\cite{sriram2002,WoodhouseGoldstein2012,Norton2018}. 
 In our entire analysis, we assume that $\mu_1$ is sufficiently small such that $\eta\nu-2\mu(\mu+\mu_1)>0$. Since we only study linear stability of the state with no order and no flow, we are justified in disregarding terms of higher order than quadratic in the order parameter. 
 Thus, $\Phi_{\alpha\beta}\approx-AQ_{\alpha\beta}$ (with $A>0$ in the isotropic phase) and Eq.~(\ref{Qeq}) takes the form
\begin{equation}
-AQ_{\alpha\beta} =2\mu E_{\alpha\beta}+\nu R_{\alpha\beta}.\label{Qeqn1}
\end{equation}

Likewise, we ignore the higher order terms in $R_{\alpha\beta}$; thus,  $R_{\alpha\beta} \approx \dot{Q}_{\alpha\beta}$. Our linearized equations are equivalent to the apolar case of the linearized equations of active matter that have appeared previously~\cite{Hatwalne2004,Kruse_etal2005,MarenduzzoOrlandiniCatesYeomans2007,LiverpoolMarchetti2006} when we set $\mu_1=0$; also we absorb a possible active term proportional to $Q_{\alpha\beta}$ in $\Phi_{\alpha\beta}$ in  Eq.~(\ref{Qeq}).

Assuming that $v_\alpha$ and $Q_{\alpha\beta}$ are proportional to $\exp( -\mathrm{i}\omega t)$, where the real part of $-\mathrm{i}\omega$ is the growth rate of the perturbations, we find using Eq.~\eqref{Qeqn1} that
\begin{equation}
Q_{\alpha\beta} =-\dfrac{2 \mu}{A-\mathrm{i}\omega\nu} E_{\alpha\beta}.\label{Qeq2}
\end{equation}
Using Eq.~\eqref{stressa}, the viscous stress $\sigma^v_{\alpha\beta}$ is given by
\begin{equation}
\sigma^v_{\alpha\beta}=2\eta_\mathrm{eff} E_{\alpha\beta},
\end{equation}
where 
\begin{eqnarray}
\eta_\mathrm{eff}&=&\dfrac{\eta A}{A-\mathrm{i}\omega\nu}\left[1-a -\dfrac{\mathrm{i}\omega\nu}{A}\left(1-\dfrac{2 \mu (\mu+\mu_1)}{\nu\eta} \right)  \right]\label{etaprime}\\
&=&\eta\frac{1-a-\mathrm{i}\omega\tau_\mathrm{lc}'}{1-\mathrm{i}\omega\tau_\mathrm{lc}}.
\end{eqnarray}
Here, we have defined the dimensionless activity by $a=a'\mu/\eta A$, and the relaxation times by $\tau_\mathrm{lc}=\nu/A$ and $\tau_\mathrm{lc}'=\tau_\mathrm{lc}[1-2\mu(\mu+\mu_1)/(\nu\eta)]$.  Since we are assuming that $\eta\nu-2\mu(\mu+\mu_1)>0$, the isotropic phase of an infinite %unbounded 
active nematic
fluid is unstable against shear flow and local ordering~\cite{swimming2018} when $a>1$.  We will see in our instability analyses for the various geometries that the critical values of the dimensionless activity correspond to a negative effective shear viscosity, i.e.  $a\ge1$. Note that in the oblate particle case of $\mu>0$, the critical value of the activity $a'$ is %always 
positive, meaning that significantly active contractile (puller) particles lead to instablity. %Whereas, 
For the prolate particle case of $\mu<0$, the critical activity is negative, meaning that sufficiently active extensile (pusher) particles lead to instability. 

It is apparent from the above equation that the effective viscosity of the fluid $\eta_\mathrm{eff}$ depends on the growth rate $-\mathrm{i}\omega$; in other words, the fluid behaves like a viscoelastic fluid due to the presence of the nematic molecules. At the special value of dimensionless activity $a=2 \mu (\mu+\mu_1)/\nu\eta=1-\tau_\mathrm{lc}'/\tau_\mathrm{lc}$, the effective viscosity $\eta_\mathrm{eff}$ is independent of $\omega$, and the fluid behaves like a Newtonian fluid with shear viscosity $\eta_\mathrm{eff}=\eta-2\mu(\mu+\mu_1)/\nu$. We will see below that at this special value of the activity the growth rate is that of a passive fluid. 

Since we ignore the inertia of the fluid, the force balance equation is given by 
\begin{equation}
\partial_\beta\sigma_{\alpha\beta}=0,\label{fb}
\end{equation}
with  $\sigma_\mathrm{\alpha\beta}=-p\delta_{\alpha\beta}+\sigma^v_{\alpha\beta}$. The pressure $p$ is the pressure arising from  the incompressibility condition $\bm{\nabla}\cdot \textbf{v}=0$. We have disregarded the Ericksen stress $\sigma^\mathrm{e}_{\alpha\beta}=\mathcal{F}\delta_{\alpha\beta}-\partial{\mathcal F}/\partial(\partial_\beta Q_{\mu\nu})\partial_\alpha Q_{\mu\nu}$~\cite{DeGennes1969,JulicherGrillSalbreaux2018} since it is at least quadratic order in $Q_{\alpha\beta}$.
Then, using the incompressibility condition $\bm{\nabla}\cdot \textbf{v}=0$,  the linearized Eq.~\eqref{fb} can be simplified to  
\begin{equation}
\eta_\mathrm{eff}\nabla^2\textbf{v}-\bm{\nabla}p=0.\label{Stokes}
\end{equation}
The incompressibility condition is imposed by representing $\textbf{v}$ as the curl of a stream function $\bm{\psi}$ i.e. $\textbf{v}=\bm{\nabla}\times\bm{\psi}$. 
For simplicity, we choose the form of $\bm{\psi}$ such that $\bm{\nabla}\cdot\bm{\psi}=0$. 
Taking the curl of Eq. \eqref{Stokes} yields
\begin{equation}
\nabla^4\bm{\psi}=0,\label{fb3} 
\end{equation}
where $\nabla^4$ is the square of the Laplacian operator in three dimensions. 
We solve the above equation with the boundary conditions appropriate to the geometry at hand and calculate the forces on the interface due to the fluid.

To describe the force per unit area acting on at the surface, we need to parametrize the surface as $\mathbf{X}(u^1,u^2)$, with coordinates $u^1$ and $u^2$. Due to the free energy associated with the surface [see Eq.~\eqref{Fi}], the force per unit area acting on the surface is given by~\cite{Powers2010}
\begin{equation}
{\mathbf f}_m=2\gamma H\mathbf{n},
\end{equation}
where $\mathbf{n}$ is the outward normal.
Note that our convention is that $H$ is negative for a sphere or a cylinder. 
Since we disregard the inertia 
 of the surface,
 the force balance equation at the surface reads
  \begin{eqnarray}
  (\sigma^+_{nn}-\sigma^-_{nn})
  +2\gamma H&=&0\\\label{forbala2}
   (\sigma^+_{n\alpha}-\sigma^-_{n\alpha})+\partial_jX^\alpha&=&0,
  \end{eqnarray}
 where $\sigma^\pm_{nn}=n^\alpha\sigma^\pm_{\alpha\beta}n^\beta$ and $\sigma_{n\alpha}=n^\beta\sigma_{\beta\alpha}$, with the plus and minus denoting the stress exerted  on the interface from the $\mathbf{n}$ and $-\mathbf{n}$ sides, respectively. 

\begin{figure*}[h]
	\includegraphics[width=%0.45\textwidth
	7in]{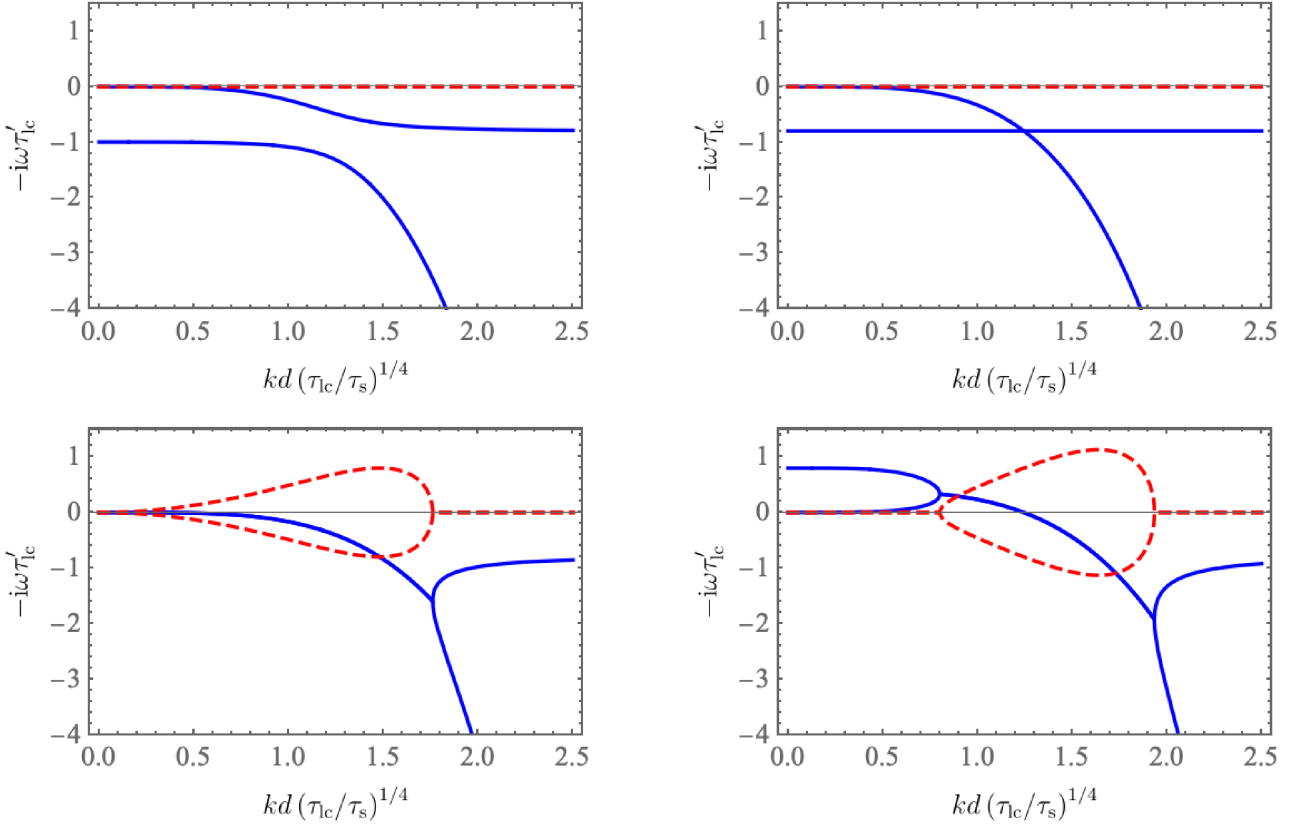}
	\centering
	\caption{(Color online.) 
		Real (blue) and imaginary (red dashed) parts of the dimensionless growth rate $-\mathrm{i}\omega \tau_\mathrm{lc}'$ of an active film of thickness $d$ as a function of dimensionless wavevector $kd$, in the limit $\tau_\mathrm{lc}/\tau_\mathrm{s}\gg 1$, for various dimensionless activities: $a=0$, corresponding to an interface of a passive liquid crystal in the isotropic phase (upper left panel); $a=0.2$ (upper right panel), corresponding to the value of activity for which the fluid behaves as a passive Newtonian fluid and the liquid crystal degrees of freedom relax independently;  $a=1$ (lower left panel) corresponding to the critical value of activity at which the system is marginally stable; and $a=1.8$, corresponding to an activity at which the system is unstable (lower right panel).  The case of $\tau_\mathrm{lc}'/\tau_\mathrm{lc}=0.8$ is shown.}
	\label{rootsplotm}
\end{figure*}

 We close this section with estimates of the  magnitudes  of the liquid crystal relaxation time $\tau_\mathrm{lc}$ and the characteristic time scales for a film with interfacial tension or bending stiffness. A crude dimensional analysis estimate for $\tau_\mathrm{lc}=\nu/A$ is to suppose $\nu\approx\eta$, and to take $A=k_\mathrm{B}T/\ell^3$, where $k_\mathrm{B}T$ is thermal energy and $\ell$ is the length of the active particles. Using the viscosity of water, $\eta\approx10^{-3}\,$N-s/m$^2$, and $\ell\approx10\,\mu$m leads to $\tau_\mathrm{lc}\approx300\,$s. If the rods are $1\,\mu$m in length, then $\tau_\mathrm{lc}\approx0.3\,$s.  However,  since we are considering an active system, it is reasonable to suppose that $A$ is not determined by thermal energy, and that $A$, and the liquid crystal relaxation rate may be much bigger. For a film of thickness $d\approx1\,$mm and for the air-water surface tension $\gamma\approx70\times10^{-3}\,$N/m, the characteristic surface-tension driven relaxation time is $\tau_\mathrm{s}=\eta d/\gamma\approx0.1\,$ms. Thus, we expect the film relaxation time to be much shorter than the liquid crystal relaxation time, and we will focus on this limit. However, due to our uncertainty about the value of $A$, and also to show some of the range of possible phenomena, we also consider the case of $\tau_\mathrm{lc}\approx\tau_\mathrm{s}$.

\section{Instability of an active fluid film}\label{membrane}

In this section, we study the instability of a flat interface  of an active nematic fluid in its isotropic phase. The fluid is a film of thickness $d$ atop a solid substrate, with air above the film (Fig.~\ref{schemf}). 

We consider an air-fluid interface with constant uniform surface tension $\gamma$, and no bending stiffness. 
 A film of passive fluid is always stable to sinusoidal perturbation, since the perturbation increases 
  the surface area. 
  Thus, the instability we study in this section arises from the activity of the fluid.

 The surface, which lies in the $zx$ plane in its unperturbed state, is subject to a transverse 
perturbation which is the real part of $h=\epsilon(t)\exp(\mathrm{i} kx)$, as shown in Fig.~\ref{schemf}. 
We assume that $\epsilon\propto\exp(-\mathrm{i}\omega t)$.

\begin{figure}[t]
\includegraphics[width=0.45\textwidth]{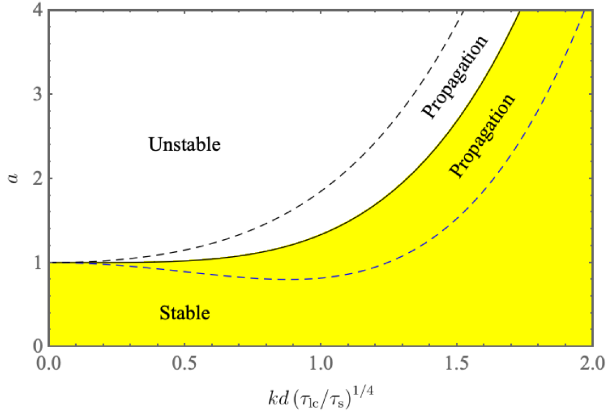}
\caption{(Color online.) Stability diagram showing when an interface of an active film is unstable as a function of dimensionless activity $a$ and dimensionless wavenumber $kd$ for the case of $\tau_\mathrm{s}\ll\tau_\mathrm{lc}$. The horizontal axis of the plot is scaled by $(\tau_\mathrm{lc}/\tau_\mathrm{s})^{1/4}$ since the longest wavelengths are unstable in this limit. The system is stable in the yellow-shaded region, and unstable in the unshaded region. Both the growing modes and the decaying modes propagate in the region between the two dashed lines.}
	\label{critact}
\end{figure}

The stream function is given by $\bm\psi=\psi\hat{\mathbf z}$, with $\psi$ a biharmonic function.  For small deflections $kh\ll1$, the kinematic condition  takes the form
\begin{equation}
v_{y}(y=0)=\partial_t h,
\end{equation}
where $v_y=-\partial_x\psi$. This condition, along with the conditions of zero tangential stress at the interface,
\begin{equation}
\sigma_{xy}(y=0)=0,
\end{equation}
and vanishing flow at $y=-d$, leads to 
\begin{eqnarray}
\psi&=&-\frac{\mathrm{i}\omega\epsilon}{k}\mathrm{e}^{\mathrm{i}kx}\left\{\left[\cosh ky+\frac{\sinh ky }{F}\right]\right.\nonumber\\
&&+\left.\left[(1-2k^2d^2F)\frac{ky \cosh ky}{F}- kx \sinh ky\right]\right\},
\end{eqnarray}
where
\begin{equation}
F=\frac{\sinh 2kd-2kd}{\cosh 2kd+2k^2d^2+1}.\label{Fofkdeqn}
\end{equation}

For small deflections the mean curvature is $H\approx-\partial^2_xh/2$, and 
 the force balance equation on the interface becomes
 \begin{equation}
 -\sigma_{yy}|_{y=0}%-\kappa \partial^4_y h
 +\gamma \partial^2_x h=0.\label{forbal}
 \end{equation}
The stress component $\sigma_{yy}$ %and $\sigma^+_{zz}$ 
can be found by  calculating the pressure from the $x$-component of the force balance equation~\eqref{fb}. 
Once $\sigma_{yy}$ is calculated, we use normal stress balance~\eqref{forbal} at $y=0$ to obtain the characteristic equation
 \begin{equation}
 -\mathrm{i}\omega=-\frac{\gamma k}{2\eta_\mathrm{eff}(\omega)}F(kd).
 \label{rootm}
 \end{equation}
 In the passive Newtonian case with $a=0$ and with no coupling between the fluid and the liquid crystalline degrees of freedom, i.e. $\mu=\mu_1=0$, the growth rate has two branches that cross, one corresponding to the negative growth rate of a Newtonian film~\cite{HenleLevine2007}, with characteristic time scale $\tau_\mathrm{s}=\eta d/\gamma$,
 \begin{eqnarray}
 -\mathrm{i}\omega&\sim&-\frac{\gamma k}{2\eta},\quad kd\gg1\\
 -\mathrm{i}\omega&\sim&-\frac{\gamma d^3k^4}{3\eta},\quad kd\ll1\label{longwavebranch}
 \end{eqnarray}
 and one corresponding to the liquid crystalline relaxation rate, $-\mathrm{i}\omega=-1/\tau_\mathrm{lc}=-A/\nu$. When $\mu$ (or $\mu_1$) is nonzero  and $a=0$, the growth rate curves repel each other instead of crossing, as in Fig.~\ref{rootsplotm}, upper left panel.
 
 The active case is like the case of a passive viscoelastic fluid~\cite{HenleLevine2007}, for which the effective shear viscosity depends on $\omega$, and we must solve Eq.~\eqref{rootm} for $\omega$ as a function of $k$, which yields
 \begin{equation}
 -\mathrm{i}\omega\tau_\mathrm{lc}'=\frac{a-1}{2}-\frac{kd F\tau_\mathrm{lc}}{4\tau_\mathrm{s}}\pm\sqrt{\left(\frac{a-1}{2}-\frac{kdF\tau_\mathrm{lc}}{4\tau_\mathrm{s}}\right)^2-\frac{kd F\tau_\mathrm{lc}'}{2\tau_\mathrm{s}}},
 \label{growthflat}
 \end{equation}
 where $F$ is given by Eq.~(\ref{Fofkdeqn}).
As the activity increases, the splitting between the two growth rate curves decreases, until the value of 
$a= 2\mu  \left(\mu +\mu _1\right)/\eta\nu$ is reached. At this special value of activity,  $\eta_\mathrm{eff}$ is independent of $\omega$, and the branches of the growth rates cross as they do in the case of $a=0$ and $\mu=0$ (Fig.~\ref{rootsplotm}, upper right panel). As the activity increases further, the real branches collapse into one branch for a range of wavevector, and the imaginary parts of the growth rate become nonzero in this same range (Fig.~\ref{rootsplotm}, lower left panel). The critical activity $a=1$ corresponds to the point at which the effective shear viscosity vanishes. When $a>1$, one of the branches of the real part of the growth rate becomes positive, and the system is unstable for  sufficiently long wavelengths (Fig.~\ref{rootsplotm}, lower right panel). 
The critical activity $a_\mathrm{c}(k)$ at which the mode $k$ is marginally stable is found by demanding that $\mathrm{Re}(-\mathrm{i}\omega)=0$: 
\begin{equation}
a_\mathrm{c}(k)=1+\frac{1}{2}kd F(kd)\tau_\mathrm{lc}/\tau_\mathrm{s}.
\end{equation}
Since $\tau_\mathrm{lc}/\tau_\mathrm{s}=(\nu\gamma)/(\eta A d)$, interfacial tension tends to suppress the instability for nonzero $k$. But even if $\tau_\mathrm{s}\ll\tau_\mathrm{lc}$, the longest wavelengths are always unstable. In this limit, the two branches of the uncoupled passive case cross when $kd\sim(\tau_\mathrm{s}/\tau_\mathrm{lc})^{1/4}$, which is why we plot the growth rates vs. $kd(\tau_\mathrm{lc}/\tau_\mathrm{s})^{1/4}$ in Fig.~\ref{rootsplotm}. The shapes of the real and imaginary parts of the growth rate curves for $\tau_\mathrm{lc}\approx\tau_\mathrm{s}$ and for $\tau_\mathrm{lc}\ll\tau_\mathrm{s}$ are qualitatively similar to the case of $\tau_\mathrm{s}\ll\tau_\mathrm{lc}$, with the main difference being that the band of unstable modes reaches further into the regime of short wavelength as $\tau_\mathrm{s}$ increases relative to $\tau_\mathrm{lc}$ (See Figs.~\ref{rootsplotm2} and \ref{rootsplotm3} in the appendix).

Note that the growth rate $-\mathrm{i}\omega$ always has an imaginary part when $a$ is sufficiently near $a_\mathrm{c}(k)$; when a mode is unstable with a sufficiently small growth rate, it also propagates. Propagating modes are found when $a_-\le a\le a_+$, where
\begin{equation}
a_\pm=1+\frac{1}{2}kdF(kd)\tau_\mathrm{lc}/\tau_\mathrm{s}\pm\sqrt{\frac{1}{2}kdF(kd)\tau_\mathrm{lc}'/\tau_\mathrm{s}}.
\label{aflatpm}
\end{equation}
Also, there are no propagating modes without the interface, since $a_+-a_-\propto\sqrt{\tau'_\mathrm{lc}/\tau_\mathrm{s}}\propto\sqrt\gamma$. 

Figure~\ref{critact} shows when the interface is stable as a function of scaled dimensionless wavenumber and dimensionless activity for the case of $\tau_\mathrm{lc}/\tau_\mathrm{s}\gg1$. The system is always stable for $a<1$; as $a$ is increased beyond $a=1$, an increasingly large band of very long wavelength modes are unstable. Growing and decaying modes  with a sufficiently small growth rate [between the dashed lines in Fig.~\ref{critact}, which are given by Eq.~(\ref{aflatpm})] are also propagating. As $\tau_\mathrm{lc}/\tau_\mathrm{s}$ decreases, the band of unstable modes is limited to shorter and shorter wavelengths. To sum up, the time scale $\tau_\mathrm{lc}$ controls the rate of growth or decay of the modes, and the time scale $\tau_\mathrm{s}$ determines which modes become unstable. Since $a_\pm$ depends on $\tau_\mathrm{s}$, the velocity of propagation $\mathrm{Re}(\omega)/k$ is determined by $d/\tau_\mathrm{s}$. 

\begin{figure}[t]
 	\includegraphics[width=%0.45\textwidth
 	3.3in]{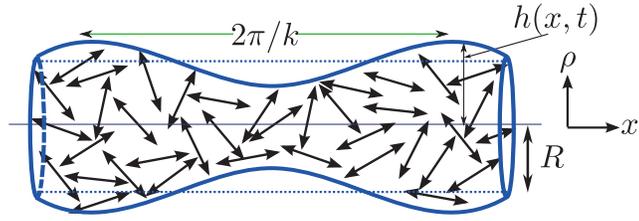}
 	\centering
 	\caption{(Color online.)
 		A cylindrical thread of isotropic active nematic liquid  subject to an axisymmetric  ripple of wavenumber $k$ and small amplitude. The double headed arrows are the active nematic molecules.}
 	\label{cylfig}
 \end{figure}
\section{Rayleigh-Plateau capillary instability}\label{thread}

A fluid thread breaks into drops because perturbations of sufficiently long wavelength lower the area of the surface, and thus the energy. This instability is known as the Rayleigh-Plateau capillary instability~\cite{Plateau1873,Rayleigh1892}.
 In this section, we study how the presence of active nematic molecules in the liquid affects the Rayleigh-Plateau capillary instability. For simplicity, we disregard the outer fluid. While this approximation was natural in our study of the stability of a flat interface between air and an active fluid, it seems less natural for a thread of active fluid, since the thread must be supported by some surrounding fluid if it is not a jet. However, unlike the passive case of a stationary cylindrical interface~\cite{Tomotika1935}, accounting for the viscosity contrast  leads to a complicated characteristic equation for the growth rate of the interface of an active thread. To avoid this complication and illustrate the essential physics, we assume the outer fluid is of sufficiently small viscosity that we may disregard it. 
   
 We consider a cylindrical fluid thread of initial radius $R$, subject to an axisymmetric harmonic perturbation of wavenumber $k$ along the $x$ direction (see Fig.~\ref{cylfig}). The cylindrical coordinates  are $(\rho,\theta,x)$. Initially the fluid is at rest, with a uniform pressure $p=\gamma/R$. The  radius of the perturbed thread is given by the real part of $h(x,t)=R+\epsilon(t) \exp(\mathrm{i} k x)$, with $\epsilon k\ll1$. For an axisymmetric flow, we follow Happel and Brenner~\cite{HappelBrenner1983} and define the stream function via $\bm{\psi}=-(\psi/\rho)\hat{\bm{\theta}}$. The stream function $\psi$ is related to velocity by $v_\rho=(1/\rho)\partial_x\psi$ and  $v_x=-(1/\rho)/\partial_\rho\psi$. % and  
 If we choose $\bm{\psi}=\Psi(\rho)\exp(\mathrm{i} k x) \hat{\bm{\theta}}$, 
 then Eq.~\eqref{fb3} in cylindrical coordinates reduces to
 \begin{equation}
 D^2\Psi=0,\label{eqpsic}
 \end{equation}
 where~\cite{Tomotika1935} $D\equiv\partial_\rho^2-(1/\rho) \partial_\rho-k^2$. 
The linearized kinematic condition at the interface,
$\bm{\nabla}\times\bm{\psi}=\partial_t{h} \hat{\bm{\rho}}$, leads to
\begin{equation}
\frac{k}{R}\Psi(\rho=R) =-\omega \epsilon.\\
\end{equation}

\begin{figure}[t]
\includegraphics[width=0.45\textwidth]{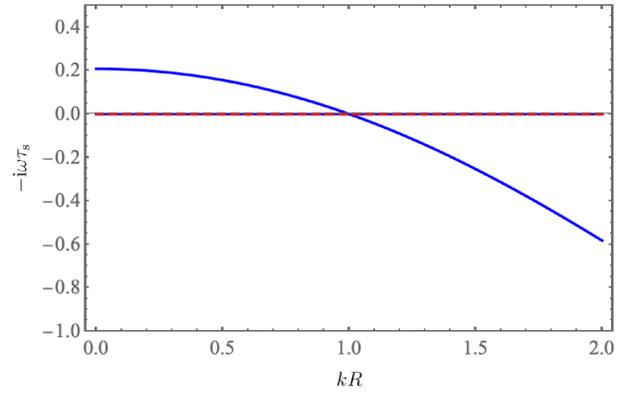}
\caption{(Color online.) Real (blue solid line) and imaginary (red dashed line) parts of the dimensionless growth rate $-\mathrm{i}\omega\tau_\mathrm{s}$ vs. dimensionless wavenumber $kR$ for $\tau_\mathrm{lc}/\tau_\mathrm{s}\gg1$. On this scale, the line corresponding the the branch $\mathrm{Re}(-\mathrm{i}\omega_-)\approx-1/\tau_\mathrm{lc}$ is along the horizontal axis. }
	\label{raycaptauslltaulc}
\end{figure}

\begin{figure*}[t]
	\includegraphics[width=%0.45\textwidth
	7in]{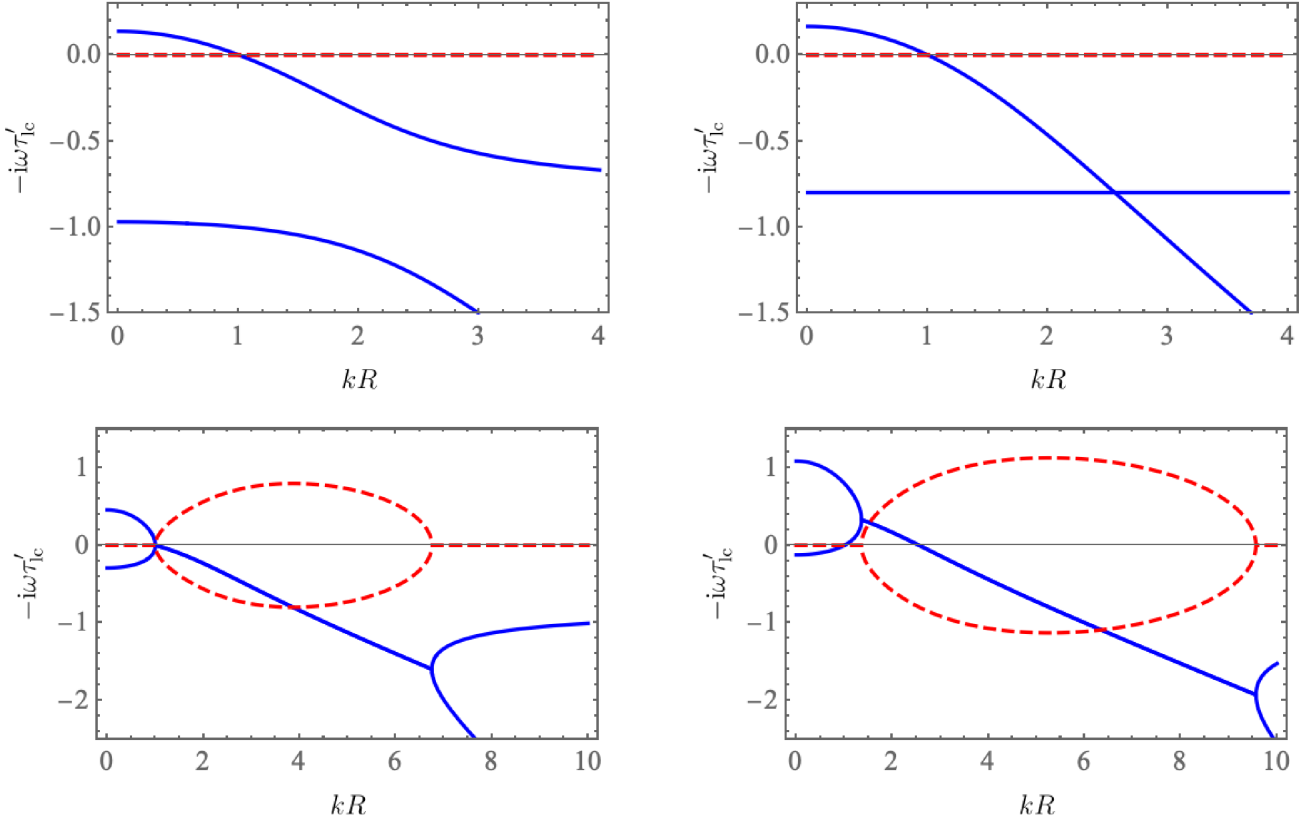}
	\centering
	\caption{(Color online.) %Schematic diagram of a  
		Real and imaginary parts of the growth rate as functions of dimensionless wavevector $kR$ for a cylindrical thread of active isotropic nematic fluid for $\tau_\mathrm{s}=\tau_\mathrm{lc}$,  $\tau'_\mathrm{lc}/\tau_\mathrm{lc}=0.8$, and dimensionless activity $a=0$ (upper left), $a=0.2$ (upper right), $a=1$ (lower left), and $a=1.8$ (upper right). 
}
	\label{cylinder-a}
\end{figure*} 
The kinematic boundary condition and the condition of zero tangential stress, $\sigma_{x\rho}|_{\rho=R}=0$, along with the condition of regularity at $\rho=0$, leads to the solution
\begin{equation}
\psi=\epsilon\omega\mathrm{e}^{\mathrm{i}kx}\left[\frac{\rho^2I_0(k\rho)}{I_1(kR)}-\frac{kRI_0(kR)+I_1(kR)}{kI_1^2(kR)}\rho I_1(k\rho)\right].
\end{equation}
where $I_0$ and $I_1$ are the  Bessel functions of first kind.
The growth rate is determined by the normal force balance equation, 
\begin{equation}
-\sigma_{\rho\rho}|_{\rho=R}+2\gamma H=0.\label{forbscyl}
\end{equation}
The pressure may be found from the $x$-component of the Stokes equation, Eq.~(\ref{Stokes}); with this pressure and the velocity field we may calculate  $\sigma_{\rho\rho}=-p+2\eta_\mathrm{eff}\partial_\rho v_\rho$ and use the mean curvature expanded~\cite{Ou-YangHelfrich} to linear order in  $\epsilon$,
 \begin{equation}
H=  -\dfrac{1}{2}\left[\frac{1}{R}+\epsilon  \left(k^2 -\frac{1}{R^2}\right)\mathrm{e}^{\mathrm{i} k x}\right],\label{surftc}
 \end{equation}
in Eq.~\eqref{forbscyl} to find
\begin{equation}
-\mathrm{i}\omega=\frac{\gamma}{2\eta_\mathrm{eff}(\omega) R}G,\label{RayPlatgrowth1}
\end{equation}
where 
\begin{equation}
G=\frac{1-k^2R^2}{k^2 R^2 I_0^2(kR)/I^2_1(kR)-(1+k^2R^2)}.
\end{equation}
When $\eta_\mathrm{eff}(\omega)=\eta$, the growth rate of Eq.~\eqref{RayPlatgrowth1} is precisely that of a thread of a passive Newtonian viscous fluid thread~\cite{Chandrasekhar1981}. Since the characteristic equation~(\ref{RayPlatgrowth1}) for the cylinder is of a similar form as the characteristic equation~(\ref{rootm}) for the planar surface, the growth rate is given by Eq.~\eqref{growthflat} with $F$ replaced by $-G/k$ and $d$ replaced by $R$. (Note that in this section $\tau_\mathrm{s}=\eta/(\gamma R)$.) Figure~\ref{raycaptauslltaulc} shows the growth  rate vs. dimensionless wavevector $kR$ for the case of $\tau_\mathrm{lc}\gg\tau_\mathrm{s}$. In this case, the growth rate is almost exactly the same as the classical result for a passive Newtonian fluid. The only dependence on activity or liquid crystalline parameters arises in the region near $kR=1$ where the real part of the growth rate vanishes. This fact can be seen by expanding the growth rate for small $\tau_\mathrm{s}/\tau_\mathrm{lc}$; away from the region where $G\ll1$, we have
\begin{eqnarray}
-\mathrm{i}\omega_-&\sim&\frac{1}{\tau_\mathrm{lc}}\\
-\mathrm{i}\omega_+&\sim&\frac{\tau_\mathrm{lc}}{\tau_\mathrm{lc}'}\frac{\gamma G}{2\eta R}.
\end{eqnarray}

\begin{figure}[t]
\includegraphics[width=0.45\textwidth]{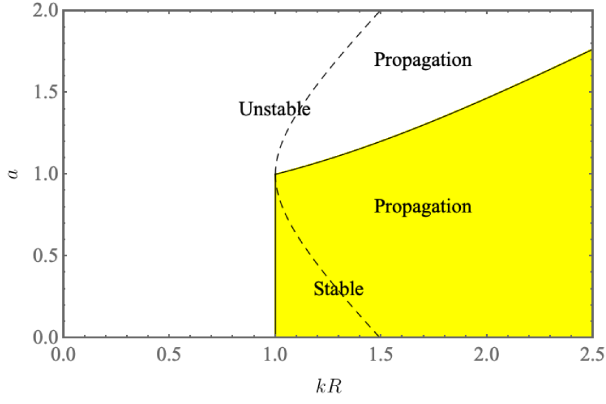}
\caption{(Color online.) Stability diagram showing when a cylindrical thread of active fluid is unstable as a function of dimensionless activity $a$ and dimensionless wavenumber $kR$ for the case of $\tau_\mathrm{s}=\tau_\mathrm{lc}$ and $\tau'_\mathrm{lc}/\tau_\mathrm{lc}=0.8$. The system is stable in the yellow-shaded region, and unstable in the unshaded region. Both the growing modes and the decaying modes propagate in the region between the two dashed lines.}
	\label{raycaptausIStaulcPhase}
\end{figure}

The effects of activity become apparent when the liquid crystal relaxation time is comparable to the film relaxation time, $\tau_\mathrm{lc}\sim\tau_\mathrm{s}$. The growth rate for several different dimensionless activities is shown in Fig.~\ref{cylinder-a}. In this case, the behavior of the growth rate with respect to activity is similar to behavior of the growth rate for a flat interface (compare with Fig.~\ref{rootsplotm}). The passive cylindrical thread is always unstable for modes with $kR<1$. Likewise, in the active case, modes with $kR<1$ are always unstable. Once $a>1$, modes with a wavenumber greater than $1/R$ can also be unstable; in particular, $\mathrm{Re}(-\mathrm{i}\omega)=0$ when
\begin{equation}
a_\mathrm{c}(k)=1-\frac{G\tau_\mathrm{lc}}{2\tau_\mathrm{s}}.
\end{equation}
Propagating modes are found when $a_-<a<a_+$,
where
\begin{equation}
a_\pm=1-\frac{G\tau_\mathrm{lc}}{2\tau_\mathrm{s}}\pm\sqrt{-\frac{2G\tau_\mathrm{lc}'}{\tau_\mathrm{s}}}.
\end{equation}
Note that propagation only occurs when $kR>1$, i.e. $G(k)<0$. Figure~\ref{raycaptausIStaulcPhase} is the stablity diagram for the case of $\tau_\mathrm{lc}=\tau_\mathrm{s}$.

\section{Instability of a spherical active droplet}\label{droplet}
\begin{figure}[t]
	\includegraphics[width=%0.45\textwidth
	3.3in]{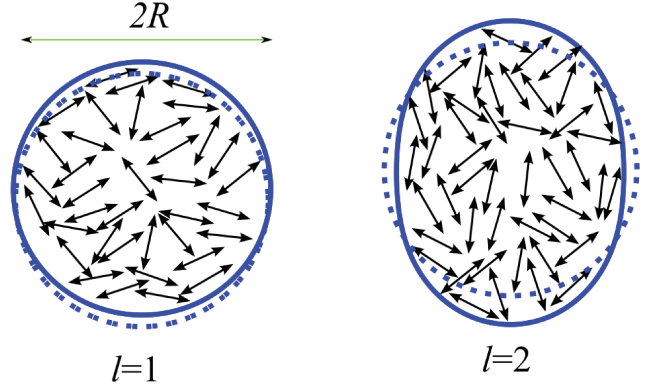}
	%\centering
	\caption{(Color online.) %Schematic diagram of a 
		A spherical droplet of isotropic active nematic liquid fluid (blue) subject to spherical harmonic ripples with $l=1$ and $l=2$. The unperturbed spherical droplet is represented by the dashed line.
		The double headed arrows are the active nematic molecules. 
	}
	\label{schems}
\end{figure}
\begin{figure}[t]
	\includegraphics[width=%0.45\textwidth
	3.3in]{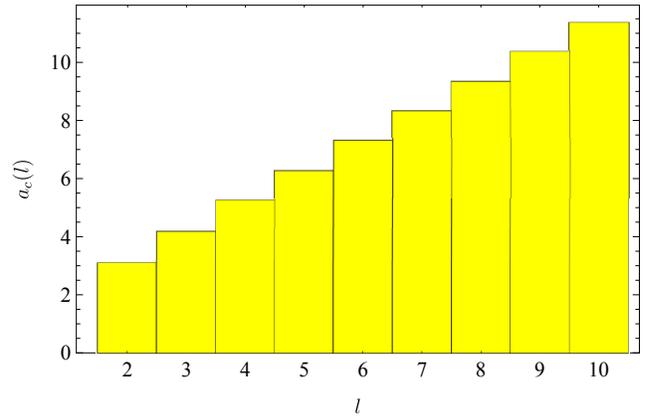}
	\centering
	\caption{(Color online.)   Dimensionless critical activity $a_c(l)$ vs $l$ for the spherical droplet at $\tau_\mathrm{lc}/\tau_\mathrm{s}=2$.
	}
	\label{spherel}
\end{figure}

A cylinder of active fluid is unstable, and breaks up into spherical droplets.
A spherical droplet of a Newtonian fluid is always stable against surface tension since the spherical shape  minimizes the surface energy. However,  a spherical  droplet of active fluid might go unstable due to activity. Here we carry out a linear stability analysis for a droplet of active nematic fluid in the isotropic phase (see Fig. \ref{schems}).  We assume that the spherical droplet of radius $R$ is subject to spherical harmonic perturbations such that the surface of the perturbed drop can be represented by $\mathbf{X}(\theta, \phi)=(R+\epsilon(t) Y^m_l(\theta, \phi))\hat{\bm{ r }}$ with $\epsilon\ll R$. 
We choose the following form of the stream function $\bm{\psi}$ to enforce the condition $\bm{\nabla}\cdot\bm{\psi}=0$:
\begin{equation}
\bm{\psi}=-v( r )\dfrac{1}{\sin\theta}\dfrac{dY^m_l(\theta, \phi)}{d\phi} \hat{\bm{\theta}}+v( r )\dfrac{dY^m_l(\theta, \phi)}{d\theta}\hat{\bm{\phi}},
\end{equation}
where  $Y^m_l(\theta, \phi)$ is a spherical harmonic. Inserting this stream function in the Stokes equations, we find that the function $v( r )$ obeys 
\begin{equation}
D^2v( r )=0,\label{eqpsis}
\end{equation} 
where 
\begin{equation}
D\equiv \dfrac{1}{ r ^2}\left[ \dfrac{d}{d r }\left( r ^2 \dfrac{d}{d r }\right) -l(l+1)\right].
\end{equation}
The boundary conditions on the interface are the linearized kinematic condition,
$\bm{\nabla}\times\bm{\psi}=\partial_t\textbf{X}$, and the linearized zero shear stress condition:
\begin{eqnarray}
-l(l+1) \dfrac{v( r =R)}{R} =-\mathrm{i}\omega\epsilon,\\
\sigma_{\phi r }( r =R)=0,\\
\sigma_{\theta r }( r =R)=0.
\end{eqnarray}
The solution of Eq. \eqref{eqpsis} with the above boundary conditions is given by 
\begin{equation}
%v( r )=\beta  \epsilon\frac{   r  ^{l-1} R^{-l-1} \left[l (l+2) R^2-\left(l^2-1\right)  r  ^2\right]}{2 l+1}.
v(r)=\mathrm{i}\omega  \epsilon\frac{  r ^l R^{-l-1} \left[l (l+2) R^2-\left(l^2-1\right) r ^2\right]}{l (l+1) (2 l+1)}
\end{equation}
With this solution, we get the following expression for $\sigma_{ r  r }$ after integrating the $ r $-component of  Eq. \eqref{fb3} with respect to $ r $:\\
\begin{equation}
\sigma_{ r  r }( r ,\theta, \phi)=-2\mathrm{i} \omega\eta_\mathrm{eff}  \epsilon \mathcal{G}[l]Y^m_l(\theta, \phi)+C,\label{strrs}
\end{equation}
where
\begin{equation}
\mathcal{G}[l]=\frac{  (l-1)     r  ^{l-2} R^{-l-1}  \left[\left(-l^3+4 l+3\right)  r  ^2+l^2 (l+2) R^2\right]}{l (2 l+1)}.
\end{equation}
In the unperturbed state, the surface tension leads to a constant pressure $C$ via the Young-Laplace law. Since we suppose that there is no fluid outside the drop,  the force balance equation at the surface of the drop  (in the limit $\epsilon\ll R$) is given by (see Eq.~\eqref{forbala2})
\begin{equation}
\sigma_{ r  r }(R,\theta, \phi)-2\gamma H=0.\label{surffs}
\end{equation}  
The mean curvature $H$ is given to first order in $\epsilon$ by \cite{Ou-YangHelfrich},
\begin{equation}
H=-\left[\frac{1}{R}+\epsilon\frac{  (l-1)(l+2) }{2R^2}Y^m_l(\theta, \phi)\right].\label{surftss}
\end{equation}
We see from Eqs. \eqref{strrs} and  \eqref{surftss} that, for the $l=1$ mode, there are no changes in  $\sigma_{ r  r }$ or the Laplace pressure $2\gamma H$ due to the perturbation, because to leading order, the $l=1$ mode is equivalent to the displacement of the droplet along the $z$ direction (see Fig. \ref{schems}). Therefore, we consider modes with $l>1$.
From Eq.  \eqref{strrs}, \eqref{surffs} and  \eqref{surftss}, we find that $C=-2\gamma/R$ and 
\begin{equation}
-\mathrm{i}\omega=-\frac{\gamma}{2\eta(\omega) R}\frac{l(l+2)(2l+1)}{2l^2+4l+3}.\label{roots}
\end{equation}
When $\eta$ is independent of $\omega$, this result is precisely the relaxation rate for perturbations of a sphere with surface tension in the limit that viscosity dominates inertia~\cite{Chandrasekhar1959,Reid1960}. 

Equation (\ref{roots}) is quadratic in $\omega$,  and the real parts of its two roots represent growth rates of the perturbation. 
The critical dimensionless activity $a_c(l)$ for the $l$th harmonic perturbation calculated  is given by
\begin{equation}
a_{c}(l)= 1+\dfrac{\tau_\mathrm{lc}}{2 \tau_\mathrm{s}}\frac{l (l+2) (2 l+1)}{2 l ^2+4l+3}.
\end{equation}
Since the smallest value of $l$ is 2, the critical value of the dimensionless activity above which droplet becomes unstable is given by
\begin{equation}
a_c(l=2)\simeq  1+\dfrac{\tau_\mathrm{lc}}{ \tau_\mathrm{s}}. 
\end{equation}
Therefore, critical dimensionless activity for a spherical droplet $a_c(l=2)$ is larger than its value for the unconfined fluid. Also, $a_c(l=2)$ decreases with $R$: smaller active droplets are more stable. Fig. \ref{spherel} shows that $a_{c}(l)$ increases almost linearly with $l$.

\section{Discussion and Conclusion}\label{conclusion}

In this paper we have studied the effect of activity on the stability of flat, cylindrical, and spherical interfaces. In all cases, the bulk instability of the active fluid, which is characterized by a vanishing effective shear viscosity, leads to spontaneous shear flows that can destabilize an interface that would be stable in the case of a passive fluid. Furthermore, all three geometries showed oscillatory behavior at suitably large activity, corresponding to propagating damped or growing modes. 
The presence of %damped or growing
propagating modes (damped or growing)  at zero Reynolds number is qualitatively different from the passive fluid case, where no propagation is seen at zero Reynolds number. The propagating modes in our linear stability analysis may be the seed for propagating modes at large amplitude, as seen in numerical calculations of active membranes~\cite{MietkeJulicherSbalzarini2019}. 
We made several approximations in this paper to make our calculation tractable. We neglected the Frank elasticity, which meant that the base state that we expanded about is uniform, $Q_{\alpha\beta}=0$. If we had included Frank elasticity, we would have to specify anchoring conditions for $Q_{\alpha\beta}$. For the case of planar or homeotropic anchoring, the base state would be nonuniform, and its stability would be more difficult to analyze by the technique we employ. The case of a zero-torque anchoring condition would lead to a uniform base state, but it would still make our calculation more complicated since we would not be able to eliminate $Q_{\alpha\beta}$ by simply solving an algebraic equation, and we would not be able to lump all the liquid-crystalline and active effects into the effective frequency-dependent viscosity $\eta_\mathrm{eff}(\omega)$. It would be interesting to generalize our calculations to include Frank elasticity, since it has been shown that Frank elasticity (or equivalently rotational diffusion in the work of Woodhouse and Goldstein) leads to spontaneous flow even for undeformed confining surfaces~\cite{WoodhouseGoldstein2012}. A second major simplification is our neglect of the outer fluid. Because we neglected the viscosity of the outer fluid, we only had to solve a quadratic equation to find the branches of the growth rate. Including the outer fluid is more realistic, and it will lead to a more complicated characteristic equation, and more branches. Also, if we use the thermal energy  scale to estimates the material parameters (questionable in a active system), we are led to $\tau_\mathrm{lc}\gg\tau_\mathrm{s}$, which makes the interesting activity-driven phenomena such as instability and oscillation occur at long wavelength in the case of the flat film, but only in a narrow regime near $kR\approx1$ in the case of the cylindrical thread. When the viscosity of the outer fluid is accounted for, the growth rate of the passive cylindrical thread vanishes~\cite{Tomotika1935} at $k=0$, which will also  lead to interesting activity-driven behavior at long wavelength in the cylinder. Finally, all of the calculations we did for interfaces could be modified to apply to the case of an active fluid bound by a membrane, which could be more relevant for biological phenomena.

\section{Appendix}\label{appen}

\label{plotappendix}
\begin{figure*}[h]
	\includegraphics[width=%0.45\textwidth
	7in]{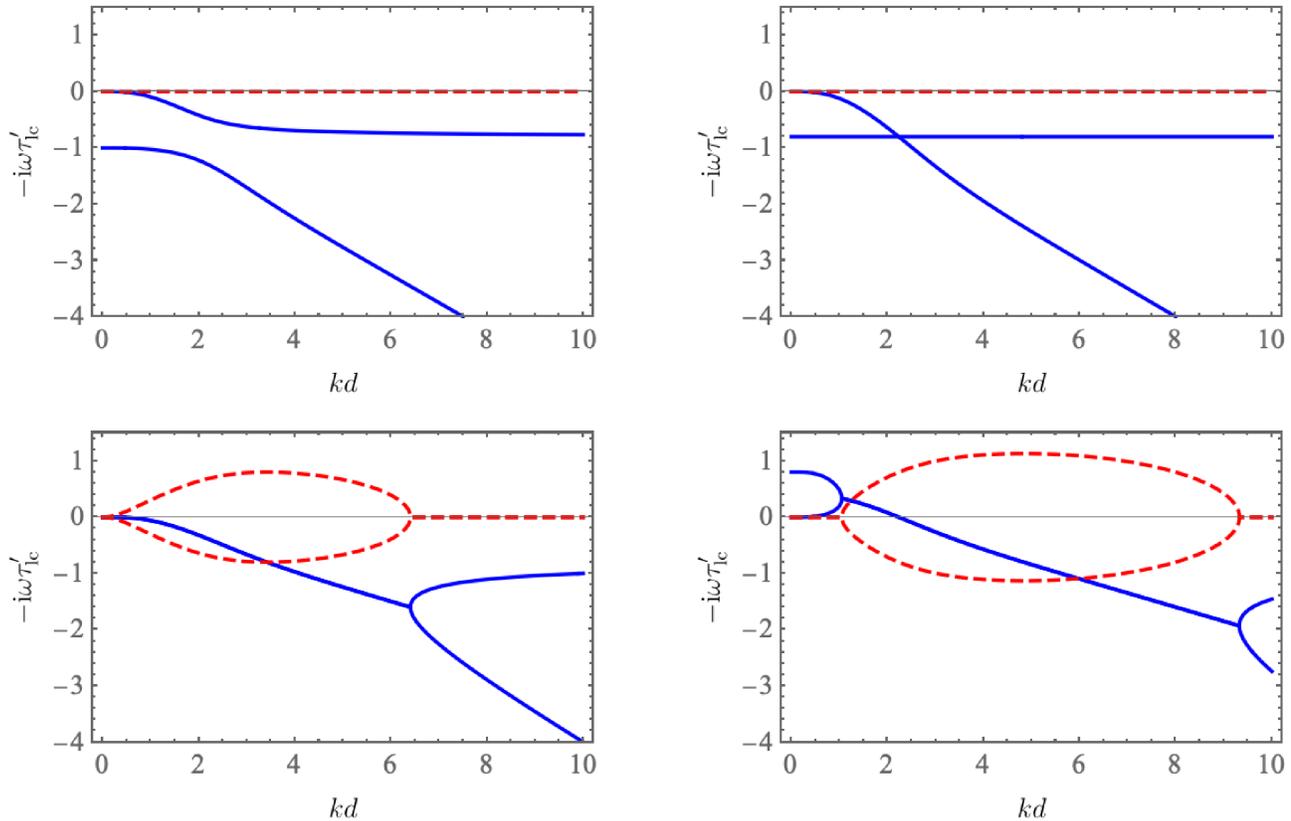}
	\centering
	\caption{(Color online.) 
		Real (blue) and imaginary (red dashed) parts of the dimensionless growth rate $-\mathrm{i}\omega \tau_\mathrm{lc}'$ of an active film of thickness $d$ as a function of dimensionless wavevector $kd$, in case $\tau_\mathrm{lc}/\tau_\mathrm{s}=1$, for various dimensionless activities: $a=0$, corresponding to an interface of a passive liquid crystal in the isotropic phase (upper left panel); $a=0.2$ (upper right panel), corresponding to the value of activity for which the fluid behaves as a passive Newtonian fluid and the liquid crystal degrees of freedom relax independently;  $a=1$ (lower left panel) corresponding to the critical value of activity at which the system is marginally stable; and $a=1.8$, corresponding to an activity at which the system is unstable (lower right panel).  The case of $\tau_\mathrm{lc}'/\tau_\mathrm{lc}=0.8$ is shown.}
	\label{rootsplotm2}
\end{figure*}

\begin{figure*}[h]
	\includegraphics[width=%0.45\textwidth
	7in]{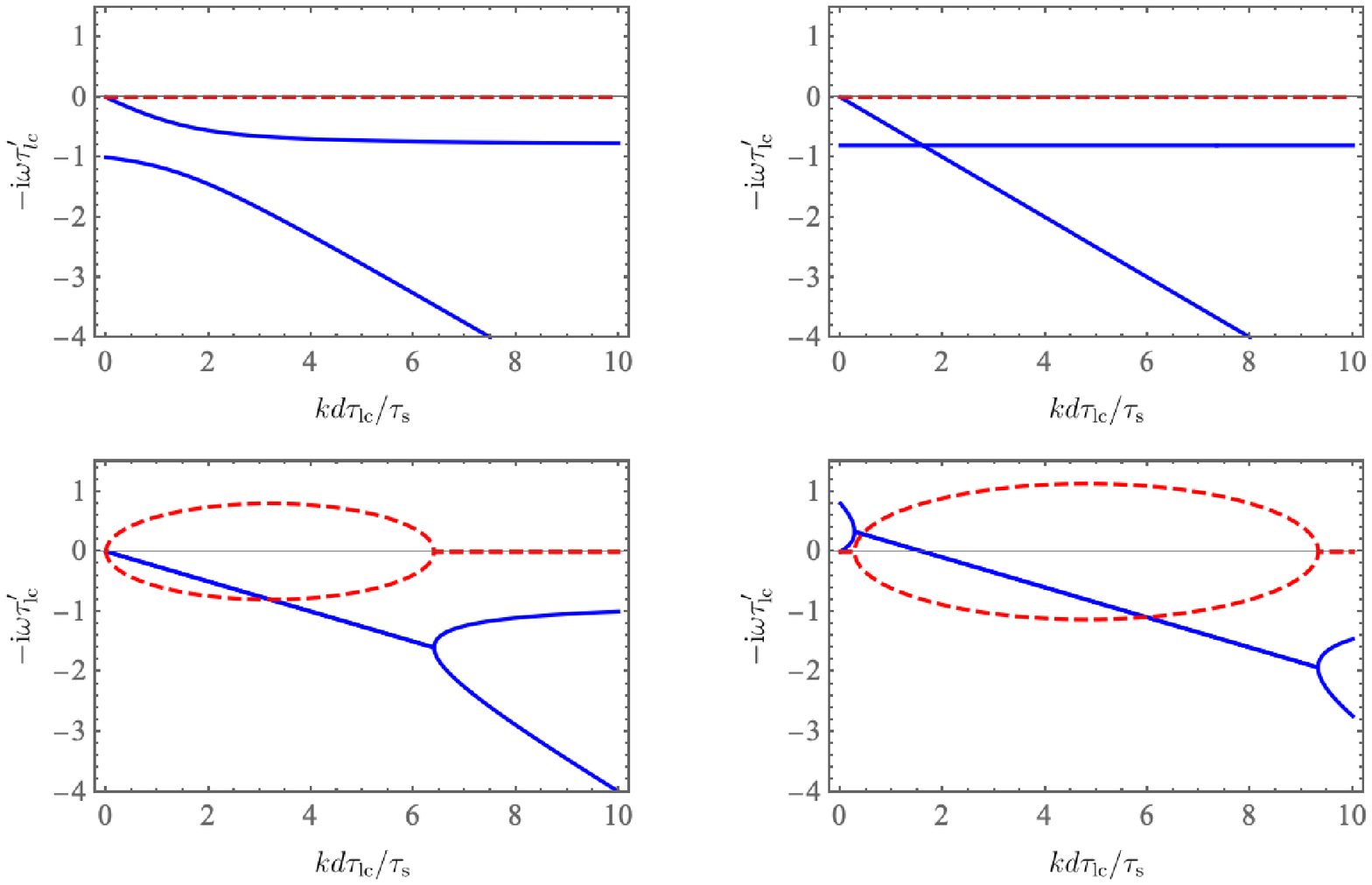}
	\centering
	\caption{(Color online.) 
		Real (blue) and imaginary (red dashed) parts of the dimensionless growth rate $-\mathrm{i}\omega \tau_\mathrm{lc}'$ of an active film of thickness $d$ as a function of dimensionless wavevector $kd$, in the limit $\tau_\mathrm{lc}/\tau_\mathrm{s}\ll 1$, for various dimensionless activities: $a=0$, corresponding to an interface of a passive liquid crystal in the isotropic phase (upper left panel); $a=0.2$ (upper right panel), corresponding to the value of activity for which the fluid behaves as a passive Newtonian fluid and the liquid crystal degrees of freedom relax independently;  $a=1$ (lower left panel) corresponding to the critical value of activity at which the system is marginally stable; and $a=1.8$, corresponding to an activity at which the system is unstable (lower right panel).  The case of $\tau_\mathrm{lc}'/\tau_\mathrm{lc}=0.8$ is shown.}
	\label{rootsplotm3}
\end{figure*}

\begin{figure}[htbp]
    \includegraphics[width=3.5in]{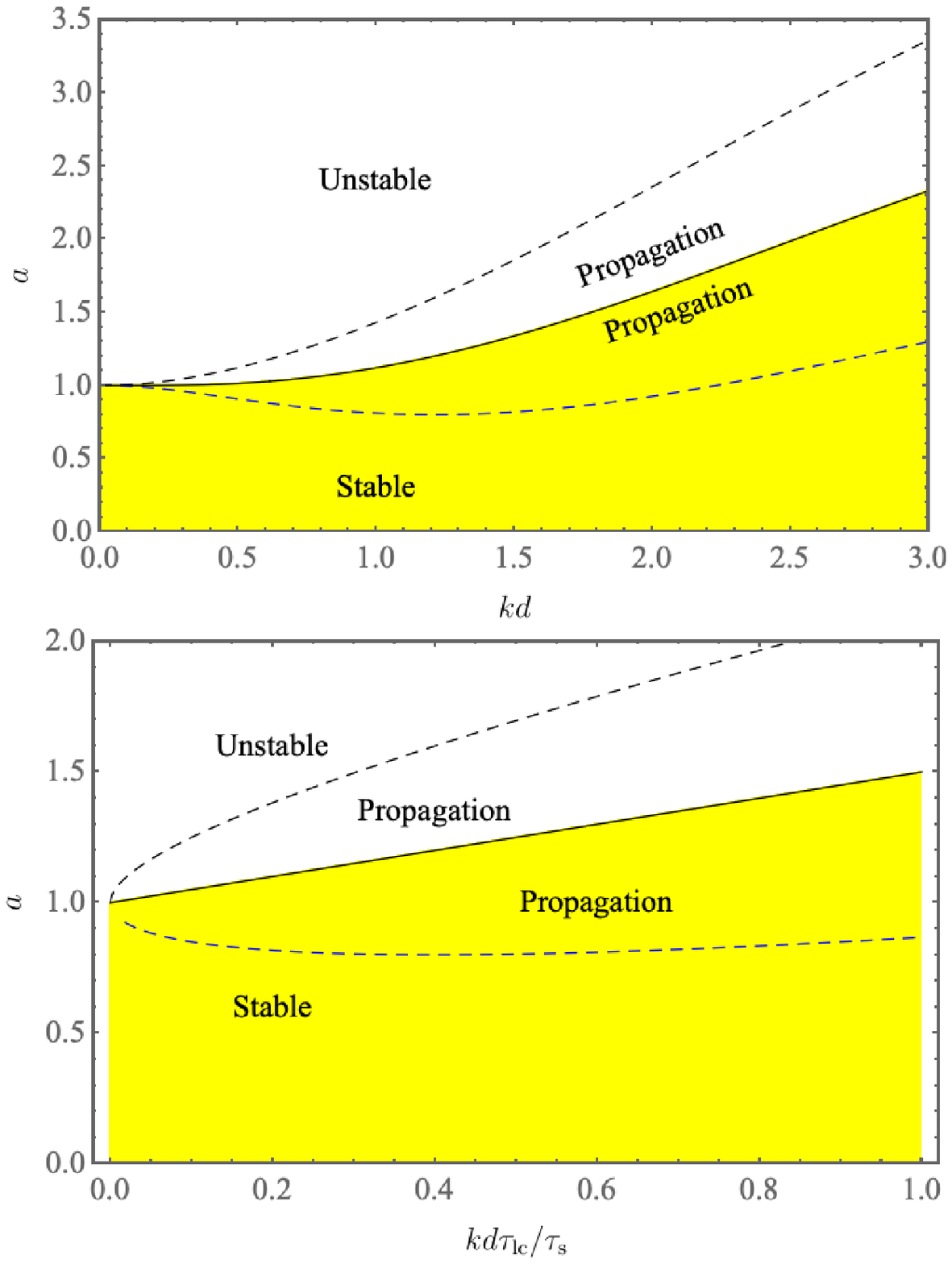}
\caption{(Color online.) Stability diagrams showing when an interface of an active film is unstable as a function of dimensionless activity $a$ and dimensionless wavenumber $kd$. The top panel shows the case of $\tau_\mathrm{s}=\tau_\mathrm{lc}$. The system is stable in the yellow-shaded region, and unstable in the unshaded region. Both the growing modes and the decaying modes propagate in the region between the two dashed lines. The bottom panel shows the case of $\tau_\mathrm{s}\gg\tau_\mathrm{lc}$, with $kd$ scaled by $\tau_\mathrm{lc}/\tau_\mathrm{s}$ since the instability occurs over a wide band of wavenumbers.}
	\label{critact2}
\end{figure}

In this appendix we display more plots of the growth rate and the stability diagram for the case of the film of thickness $d$ (Section \ref{membrane}). Fig.~\ref{rootsplotm2} shows the real and imaginary parts of the growth rate for $\tau_\mathrm{lc}=\tau_\mathrm{s}$, whereas Fig.~\ref{rootsplotm3} shows the same quantities for the case of $\tau_\mathrm{s}/\tau_\mathrm{lc}\gg1$. In all case, the shape of the curves is qualitatively similar, but the scale of wavevectors where the instability and oscillations changes, with the instability and oscillations occurring when $kd\sim(\tau_\mathrm{s}/\tau_\mathrm{lc})^{1/4}$ when $\tau_\mathrm{s}/\tau_\mathrm{lc}\ll1$, when $kd\sim1$ when $\tau_\mathrm{s}/\tau_\mathrm{lc}\sim1$, and when $kd\sim\tau_\mathrm{s}/\tau_\mathrm{lc}$ when $\tau_\mathrm{s}/\tau_\mathrm{lc}\gg1$. Figure~\ref{critact2} shows the stability diagram for $\tau_\mathrm{s}=\tau_\mathrm{lc}$  (upper panel) and $\tau_\mathrm{s}\gg\tau_\mathrm{lc}$ (lower panel).

\section{Conflicts of interest} There are no conflicts to declare. %{\bf According to the authors' guide we need this section}

\section*{Acknowledgements}

This work was supported in part by National Science Foundation Grant Nos. CBET-1437195 (TRP) and National Science Foundation Grant No. MRSEC-1420382 (RAP and TRP). We are grateful to Dan Blair, Kenny Breuer, and Ian Wong for helpful discussions.

\footnotesize{
\providecommand*{\mcitethebibliography}{\thebibliography}
\csname @ifundefined\endcsname{endmcitethebibliography}
{\let\endmcitethebibliography\endthebibliography}{}

\bibliographystyle{rsc} %the RSC's .bst file
}

%\end{multicols}
\end{document}